\documentclass[]{article}

\usepackage{graphics}
\usepackage{graphicx}
\usepackage[linesnumbered,ruled,vlined]{algorithm2e}
\usepackage{algorithmicx}
\usepackage{algpseudocode}
\usepackage[]{algorithm2e}
\usepackage{amsfonts}
\usepackage{amsmath}
\usepackage{bm}
\usepackage{bbm}
\usepackage{array}
\usepackage{breqn}
\DeclareMathOperator*{\argmax}{\arg\!\max}

\usepackage{epstopdf}
\usepackage{booktabs}
\usepackage{natbib}
\usepackage{siunitx}

\usepackage{adjustbox}
\usepackage{multirow}
\usepackage{color}

\graphicspath{{./Figures/}}

\newcolumntype{d}{>{\displaystyle}c}

\definecolor{red}{rgb}{0, 0, 0}
\usepackage{subfigure}

\graphicspath{{./Figures/}}

\begin{document}



\title{Maximization of the degree of polarization to estimate the polarization orientation angle from PolSAR data: An insight into the real and complex rotation of the coherency matrix}
\author{Avik Bhattacharya\\ Centre of Studies in Resources Engineering (CSRE),\\Indian Institute of Technology Bombay, Powai-400076, India.
}
\date{}

\maketitle

\begin{abstract}
The angle of rotation of any target about the radar line of sight (LOS) is known as the polarization orientation angle. The orientation angle is found to be non-zero for undulating terrains and man-made targets oriented away from the radar LOS. This effect is more pronounced at lower frequencies (eg. L- and P- bands). The orientation angle shift is not only induced by azimuthal slope but also by range slope. This shift increases the cross-polarization (HV) intensity and subsequently the covariance or the coherency matrix becomes reflection asymmetric. Compensating this orientation angle prior to any model-based decomposition technique for geophysical parameter estimation or classification is crucial. In this paper a new method is proposed to estimate the orientation angle based on the maximization of the degree of polarization. The proposed method is then used to infer the change in the degree of polarization with the associated orientation angle.  
\end{abstract}
\section{Introduction}
There are two main effects of surface slope and oriented urban structures on synthetic aperture radar (SAR) signals. One effect is related to the change in the radiometric property and the other is related to the change in the polarization states of the backscattered signal. The angle of rotation of any target about the radar line of sight (LOS) is known as the polarization orientation angle, $\theta$. In general, the basic principle of orientation compensation is to rotate the data about the LOS by $-\theta$. This process is analogous to rotating the antenna basis about the LOS by an angle $\theta$ such that the cross-polarized return (HV) is minimum. It has been observed that the orientation is non zero for undulating terrains and man-made targets tilted away from the radar LOS. These shifts of the orientation angle from zero are more pronounced in the low frequency (eg. L- and P- band) polarimetric synthetic aperture radar (PolSAR) data. It has been shown in~\citep{Kimura08} that the orientation shift is not only induced by terrain slope but also by artificial structures in urban area tilted away from radar LOS. Compensating this orientation prior to any model-based decomposition~\citep{freeman98}~\citep{YAMAGUCHI2011}~\citep{singh13} technique for geophysical parameter estimation or classification is very crucial. The decomposition methods proposed by Yamaguchi~\citep{Yamaguchi05}, An~\citep{Wentao10} and Lee and Ainsworth~\citep{Lee2011} reduced the number of independent parameters of the coherency matrix $\mathbf{[T]}$ from nine to eight by real rotating the coherency matrix. The decomposition yielded better results than before by using six out of eight parameters. The real and imaginary parts of $(T_{13})$ are unaccounted for in all these decomposition. The general four-component decomposition proposed by Singh~\citep{singh13} uses a complex unitary transformation to the already real unitary rotated coherency matrix. This transformation completely eliminate the $(T_{23})$ element and accounts for seven out of seven independent parameters of the coherency matrix. In a similar way as before, the complex orientation angle is estimated by minimizing the cross-polarized term $(T_{33}(\theta))$. On the other hand, the eigenvector-eigenvalue decomposition parameters like the average scattering mechanism ($\alpha$), the entropy ($H$), the anisotropy ($A$) of the Cloude-Pottier decomposition~\citep{CLOUDE97} and the scattering mechanism ($\alpha_{s}$), the scattering mechanism phase ($\phi_{\alpha_{s}}$) and the helicity ($\tau_{m}$) of the Touzi decomposition~\citep{TOUZI2007} are all roll invariant. 

The orientation angle estimation methods can be broadly categorized into two groups: (1) orientation angle derived from Digital Elevation Model (DEM) and (2) orientation angle derived from PolSAR data. The DEM obtained from SAR interferometry can also be used to compensate PolSAR data. The slope and the azimuth estimated from the DEM can then be used to obtain the orientation angle. Apart from the DEM derived orientation angle there are few other methods available in the literature which directly uses PolSAR data to compute the orientation angle. The co-polarized peak shift in polarization signature proposed in~\citep{Schuler96} is used to estimate the orientation angle in the range $[-\frac{\pi}{4},\frac{\pi}{4}]$. The phase difference between the RR-LL (Right-Right and Left-Left) circular polarizations has been used in~\citep{Lee2000} to estimate the orientation angle. The angle is in the range of $[-\frac{\pi}{4},\frac{\pi}{4}]$ and the method is computationally simpler than the co-polarized peak shift method. The minimization of the cross-pol power to estimate the orientation angle has been proposed in~\citep{Xu05}. The estimated orientation angle is dependent on $\mbox{Re}(T_{23})$, $T_{22}$ and $T_{33}$ elements of the 3$\times$3 coherency matrix, where $\mbox{Re}(X)$ denotes the real part of $X$. It has been shown that the cross-pol power minimization produces the same results as the circular polarization method. The orientation angle can also be estimated from the Cloude-Pottier eigenvalue/eigenvector decomposition~\citep{Cloude96}. The above mentioned methods have been widely used to compensate the polarization orientation effects prior to any geophysical parameter extraction or classification. 

In the following sections we will look into the estimation of the real (coherency matrix rotated by a real unitary matrix) and complex (real rotated coherency matrix rotated by a complex unitary matrix) orientation angles by maximizing the degree of polarization with an insight into the change of the degree of polarization with the associated orientation angles.

\section{Methodology}
\subsection{Real orientation angle}
To estimate the real orientation angle, the multi-looked Hermitian positive semi-definite 3$\times$3 coherency matrix $\mathbf{[T]}$ which is obtained from the averaged outer product of the target vector with its conjugate is rotated by a real unitary matrix $\mathbf{[U_{3R}]}$ given as,
\begin{equation}
\begin{split}
\mathbf{[T(\theta)]} = \mathbf{[U_{3R}]}&\mathbf{[T]}{\mathbf{[U_{3R}]^{-1}}} \\ \\
\mathbf{[T]} = \left[ \begin{array}{ccc}
T_{11} &\quad T_{12} &\quad T_{13} \\
T_{12}^{*} &\quad T_{22} &\quad T_{23} \\
T_{13}^{*} &\quad T_{23}^{*} &\quad T_{33}
\end{array}\right] ; \qquad &
\mathbf{[U_{3R}]} = \left[ \begin{array}{ccc}
1 &\quad 0 &\quad 0 \\
0 &\quad \cos(2\theta) &\quad \sin(2\theta) \\
0 &\quad -\sin(2\theta) &\quad \cos(2\theta)
\end{array}\right]
\end{split}
\label{eq:orienting_T_matrix}
\end{equation}
where $\theta \in [-\frac{\pi}{4},\frac{\pi}{4}]$. In~\citep{Lee2011}, the orientation angle, $\theta$ is estimated by minimizing the cross-polarization $(\mbox{HV})$ response given in equation~\eqref{eq:lee_angle}. The effect of $\theta$ on the three diagonal elements of the coherency matrix shows that: (1) $T_{11}=\left|\mbox{HH} + \mbox{VV}\right|^{2}/2$ is roll invariant for any $\theta$, (2)  $T_{22}=\left|\mbox{HH} - \mbox{VV}\right|^{2}/2$ always increases or remains the same after the orientation angle compensation, (3) $T_{33}=2\left|\mbox{HV}\right|^{2}$ always decreases or remains the same after orientation angle compensation.
\begin{equation}
\theta = \frac{1}{4}\tan^{-1}\left(\frac{-2\mbox{Re}(T_{23})}{T_{33}-T_{22}}\right)
\label{eq:lee_angle}
\end{equation}

\begin{figure}[h!]
\centering
\subfigure[]{\includegraphics[width=0.4\textwidth]{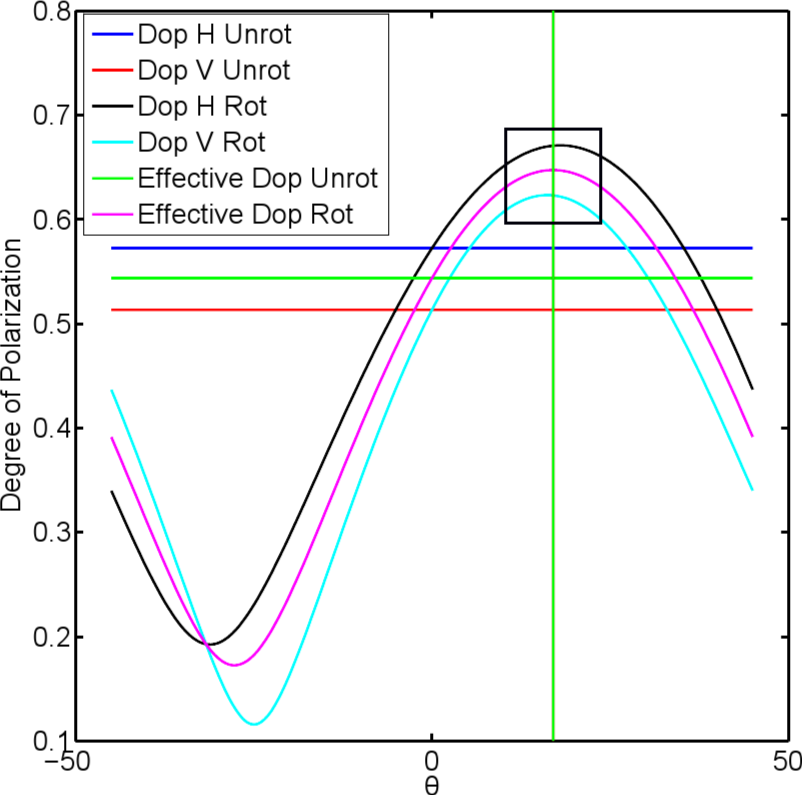}} 
\subfigure[]{\includegraphics[width=0.4\textwidth]{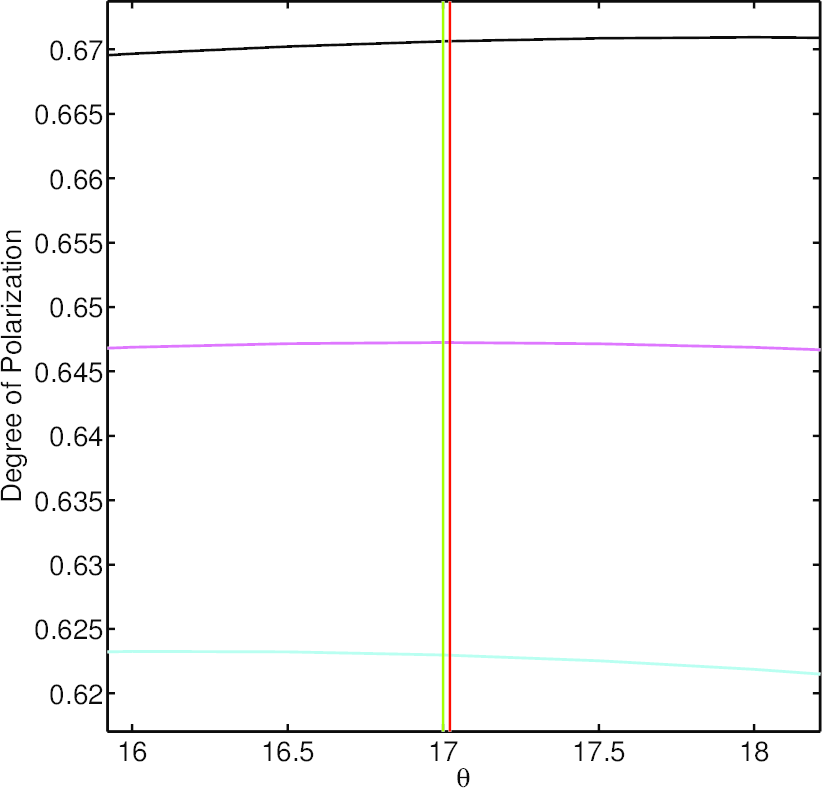}}\caption{(a). The variation of $p_{H}$, $p_{V}$ and $p_{E}$ with $\theta$, (b). Zoomed area of the square in (a) showing the similarity in the estimation of the orientation angle by the two methods.}
\label{fig:theory_results_R}
\end{figure}

Besides, multi-look PolSAR data in general can be represented by a 4$\times$4 Mueller matrix $\mathbf{[M]}$, which can be directly deduced from a 4$\times$4 coherency matrix $\mathbf{[T]}$. Since our interest centers on the special case of backscatter (BSA convention), the fourth row and column becomes zero due to reciprocity ($S_{\mbox{HV}}=S_{\mbox{VH}})$. In this case, $\mathbf{[T]}$ reduces to a 3$\times$3 coherency matrix, although $\mathbf{[M]}$ is a real 4$\times$4 matrix. 
The Mueller matrix is a linear mapping between the input and the output Stokes vectors. The received Stokes vector $\mathbf{G_{H}^{r}(\theta)} = [g_{H1}^{r} \quad g_{H2}^{r} \quad g_{H3}^{r} \quad g_{H4}^{r}]^{T}$ for a linear horizontally (H) polarized EM wave on transmit and the received Stokes vector $\mathbf{G_{V}^{r}(\theta)} = [g_{V1}^{r} \quad g_{V2}^{r} \quad g_{V3}^{r} \quad g_{V4}^{r}]^{T}$ for a linear vertically (V) polarized EM wave  on transmit are related by the Mueller matrix as shown in equation~\eqref{eq:stokes_transmit_receive},
\begin{equation}
\begin{split}
\mathbf{G_{H}^{r}(\theta)} &=  \mathbf{[M(\theta)]}\mathbf{G_{H}^{t}} \\ 
\mathbf{G_{V}^{r}(\theta)} &=  \mathbf{[M(\theta)]}\mathbf{G_{V}^{t}}
\end{split}
\label{eq:stokes_transmit_receive}
\end{equation}
where $\mathbf{G_{H}^{t}} = [1 \quad 1 \quad 0 \quad 0]^{T}$ and $\mathbf{G_{V}^{t}} = [1 \quad {-1} \quad 0 \quad 0]^{T}$ are the transmitted linear horizontal and vertical polarized Stokes vectors respectively and $\mathbf{[M(\theta)]}$ is the rotated Mueller matrix derived from $\mathbf{[T(\theta)]}$. Here, the superscript $T$ denotes the vector transpose. The state of polarization of an EM wave is characterized in terms of the degree of polarization $(0\le p\le 1)$. The degree of polarization is defined as the ratio of the (average) intensity of the polarized portion of the wave to that of the (average) total intensity of the wave. For a completely polarized EM wave, $p=1$ and for a completely unpolarized EM wave, $p=0$. In between these two extreme cases, the EM wave is said to be partially polarized, $0\le p \le 1$. The degree of polarization of a received EM wave for a horizontally and a vertically transmitted wave is defined as $p_{H}(\theta)$ and $p_{V}(\theta)$ respectively as given in equation~\eqref{eq:dop_h_v}.
\begin{equation}
p_{H}(\theta) = \frac{\sqrt{(g_{H2}^{r})^{2}+(g_{H3}^{r})^{2}+(g_{H4}^{r})^{2}}}{g_{H1}^{r}} \quad;\qquad
p_{V}(\theta) = \frac{\sqrt{(g_{V2}^{r})^{2}+(g_{V3}^{r})^{2}+(g_{V4}^{r})^{2}}}{g_{V1}^{r}}, 
\label{eq:dop_h_v}
\end{equation}
where the effective degree of polarization $p_{E}(\theta)$ is defined as, 
\begin{equation}
p_{E}(\theta) = \sqrt{\frac{p_{H}^{2}(\theta)+p_{V}^{2}(\theta)}{2}}.
\label{eq:effective_dop}
\end{equation}
The orientation angle $\theta$ is estimated by maximizing this effective degree of polarization $p_{E}(\theta)$ in the range of $\left[-\frac{\pi}{4}, \frac{\pi}{4}\right]$ as,
\begin{equation}
\theta = \argmax_{-\pi/4\le\theta\le \pi/4}\left\{ p_{E}(\theta)\right\}.
\label{eq:real_oa}
\end{equation}
Finally this estimated orientation angle is restricted in the range $\left[-\frac{\pi}{8}, \frac{\pi}{8}\right]$ by the procedure given in equation~\eqref{eq:final_OA} so as to compare it with the orientation angle estimated by the method given in~\citep{Lee2011} which lies in the range $\left[-\frac{\pi}{8}, \frac{\pi}{8}\right]$,
\begin{equation}
\theta_{0}=
\begin{cases}
\theta+\pi/4, & \text{if}\  \theta<-\pi/8 \\
\theta-\pi/4, & \text{if}\  \theta>\pi/8 \\
\theta, & \text{otherwise}.
\end{cases}
\label{eq:final_OA}
\end{equation}
To demonstrate the proposed methodology we have considered a coherency matrix $\mathbf{[T]}$ from a rotated urban area as an example, 
\begin{equation}
\mathbf{[T]} = \left[ \begin{array}{ccc}
23.66 &\qquad 2.46 + 0.61i &\qquad -0.01 - 2.03i \\
2.46 - 0.61i &\qquad 20.58 &\qquad 6.74 - 0.06i \\
-0.01 + 2.03i &\qquad 6.74 + 0.06i &\qquad 15.15
\end{array}\right].
\label{eq:T_matrix_example}
\end{equation}
The real orientation angle estimated from the proposed method is, $\theta_{0}=17^\circ$, which is similar to the one estimated by the method proposed in~\citep{Lee2011}. The variation of $p_{H}$, $p_{V}$ and $p_{E}$ with $\theta$ is shown in Figure~\ref{fig:theory_results_R}(a) and the vertical green and red lines in the zoomed Figure~\ref{fig:theory_results_R}(b) shows the estimated angles from the proposed and the method in~\citep{Lee2011} respectively.  

\subsection{Complex orientation angle}
The complex orientation angle has been primarily used in the generalized four component decomposed proposed by~\citep{singh13} to completely eliminate the $T_{23}$ component which is responsible for helical scattering. By eliminating the $T_{23}$ component of the coherency matrix, the number of independent information becomes seven for which the proposed decomposition performs better than the three component Freeman and Durden~\citep{freeman98} decomposition and the four component scattering power decomposition with real rotation of the coherency matrix by Yamaguchi \emph{et~al.}~\citep{YAMAGUCHI2011}. The idea of complex orientation is to rotate an already real rotated $\mathbf{[T(\theta)]}$ matrix by a complex unitary matrix $\mathbf{[U_{3C}]}$ given as,
\begin{eqnarray}
\mathbf{[T(\phi)]} & = & \mathbf{[U_{3C}]}\mathbf{[T(\theta_{0})]}{\mathbf{[U_{3C}]^{-1}}}\\
\mathbf{[T(\theta_{0})]} & = & \left[ \begin{array}{ccc}
T_{11}(\theta_{0}) &\quad T_{12}(\theta_{0}) &\quad T_{13}(\theta_{0}) \\
T_{12}^{*}(\theta_{0}) &\quad T_{22}(\theta_{0}) &\quad T_{23}(\theta_{0}) \\
T_{13}^{*}(\theta_{0}) &\quad T_{23}^{*}(\theta_{0}) &\quad T_{33}(\theta_{0})
\end{array}\right]\\
\mathbf{[U_{3C}]} & = & \left[ \begin{array}{ccc}
1 &\quad 0 &\quad 0 \\
0 &\quad \cos(2\phi) &\quad j\sin(2\phi) \\
0 &\quad j\sin(2\phi) &\quad \cos(2\phi)
\end{array}\right]
\label{eq:orienting_T_matrix_complex}
\end{eqnarray}
where $\phi \in [-\frac{\pi}{4},\frac{\pi}{4}]$. Unlike before, where the real part of the $T_{23}$ element ($\mbox{Re}(T_{23})$) was used to compute the real orientation angle, the complex orientation angle is estimated by using the imaginary part of the $T_{23}(\theta_{0})$ element $(\mbox{Im}(T_{23}(\theta_{0})))$ instead as shown below, 

\begin{equation}
\phi = \frac{1}{4}\tan^{-1}\left(\frac{-2\mbox{Im}(T_{23}(\theta_{0}))}{T_{33}(\theta_{0})-T_{22}(\theta_{0})}\right).
\label{eq:lee_angle_complex}
\end{equation}
Here, the effective degree of polarization is a function of complex orientation angle and as before, it is obtained from the horizontally and vertically polarized transmitted EM wave for which the degree of polarization are $p_{H}(\phi)$ and $p_{V}(\phi)$ respectively using the similar expression as in~\eqref{eq:effective_dop}. Further, following the similar procedure used for the estimation of real orientation angle, the complex orientation angle is estimated by maximizing the effective degree of polarization $p_{E}(\phi)$ in the range $\left[-\frac{\pi}{4}, \frac{\pi}{4}\right]$. Now, to compare the estimated angle $\phi$ with the one obtained in~\citep{singh13}, it is restricted in the range $\left[-\frac{\pi}{8}, \frac{\pi}{8}\right]$ using~\eqref{eq:final_OA} and the complex orientation angle lying in this range is denoted by $\phi_{0}$.

We have again considered the same $\mathbf{[T]}$ matrix given in~\eqref{eq:T_matrix_example} and rotated it with the estimated real orientation angle $(\theta_{0}=17^\circ)$. The real rotated coherency matrix, $\mathbf{[T(\theta_{0})]}$ is then subsequently unitary rotated by $\mathbf{[U_{3C}]}$. The complex orientation angle estimated  from the proposed method is, $\phi_{0}=-0.11^\circ$, which is similar to the one estimated by the method proposed in~\citep{singh13}.
\begin{figure}[h!]
\centering
\subfigure[]{\includegraphics[width=0.4\textwidth]{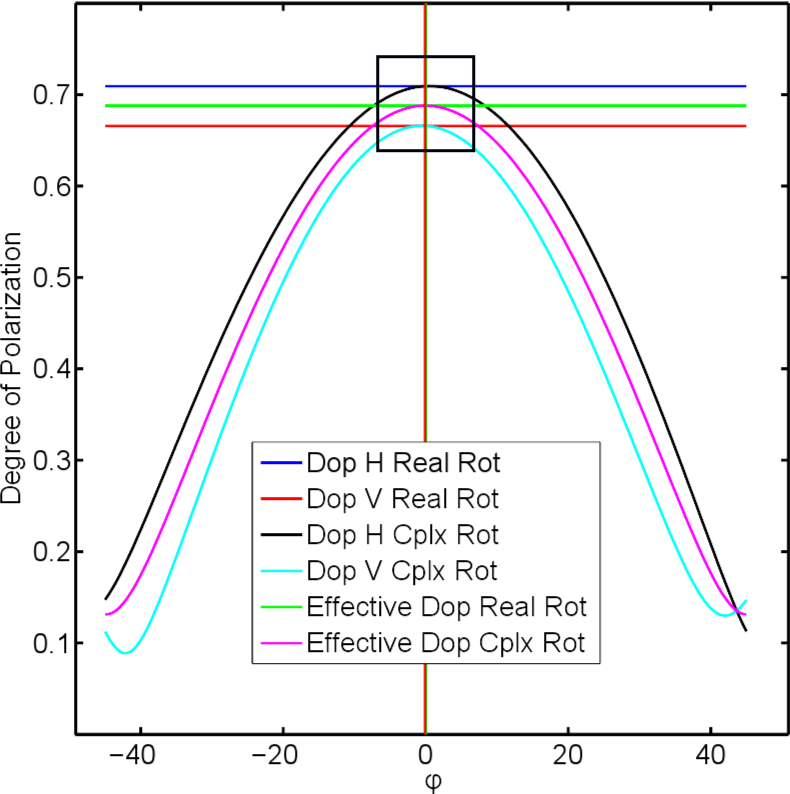}} 
\subfigure[]{\includegraphics[width=0.4\textwidth]{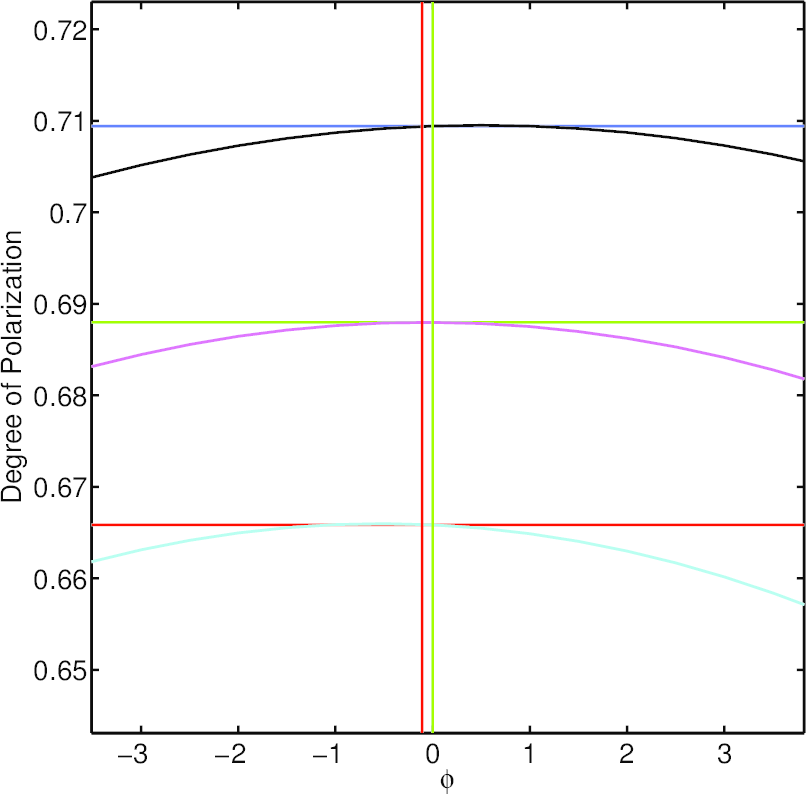}}
\caption{(a). The variation of $p_{H}$, $p_{V}$ and $p_{E}$ with $\phi$, (b). Zoomed area of the square in (a) showing the similarity in the estimation of the orientation angle by the two methods.}
\label{fig:theory_results_C}
\end{figure}
The variation of $p_{H}$, $p_{V}$ and $p_{E}$ with $\phi$ is shown in Figure~\ref{fig:theory_results_C}(a) and the vertical green and red lines  in the zoomed Figure~\ref{fig:theory_results_C}(b) shows the estimated angles from the proposed and the method in~\citep{singh13} respectively.  

\section{Study area and dataset}

For this study, a subset of an UAVSAR image acquired over Hayward, California, USA has been chosen. The sensor was operated in L-Band, fully polarimetric mode with a bandwidth of 80 MHz capable of delivering a pixel spacing of 0.6m x 1.6m in Single Look Complex (SLC) mode. The multilooked product which is used in this study has 3 looks in range and 12 looks in azimuth applied to it to produce an image of resolution 5m x 7.2m. The extracted scene has a center latitude of $37^\circ46'01.65"$N and longitude of $122^\circ12'39.78"$W. 
\begin{figure}[h!]
\centering
\subfigure[]{\includegraphics[width=0.495\textwidth]{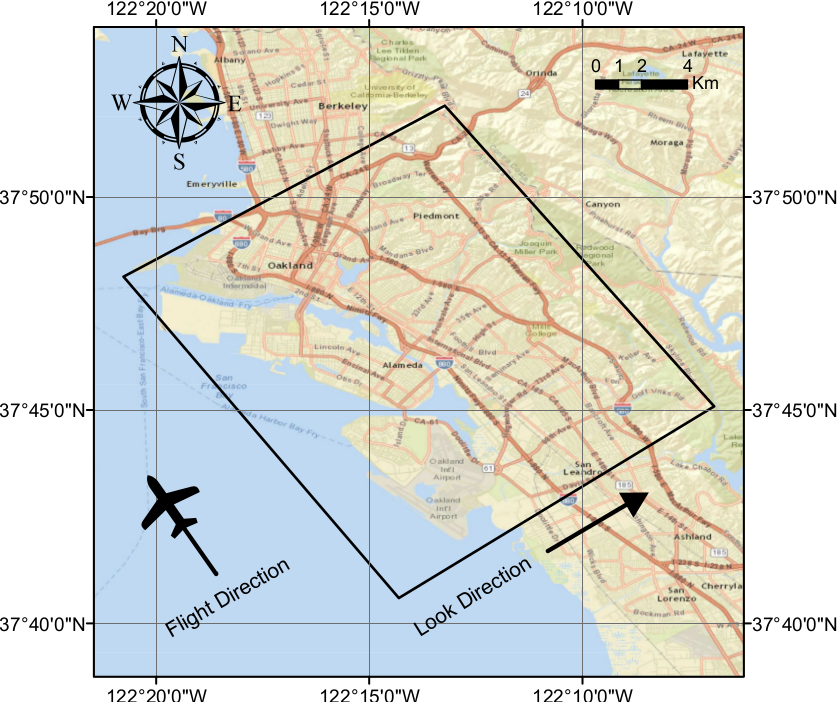}} \hspace{5mm}
\subfigure[]{\includegraphics[width=0.31\textwidth]{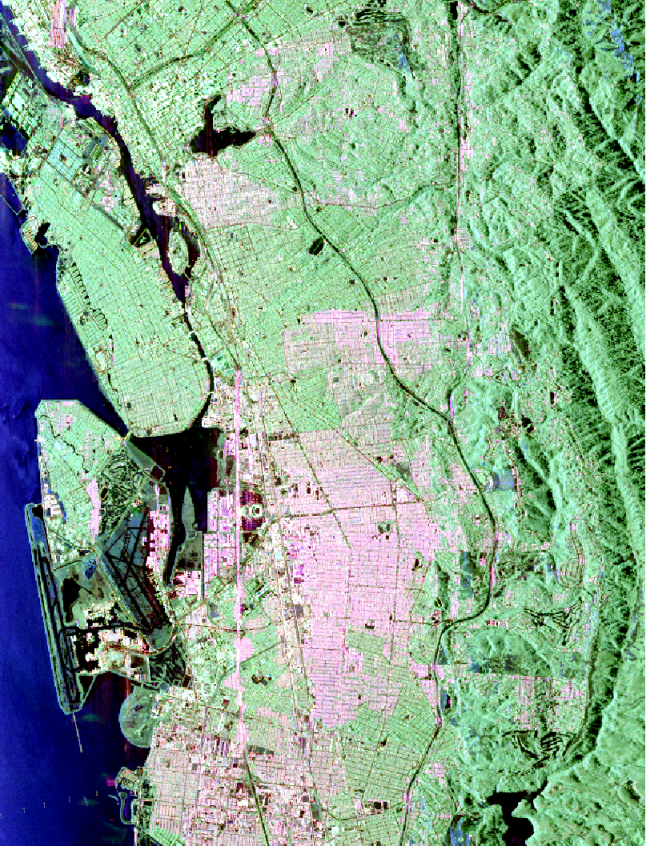}}
\caption{(a) Street map showing the extent of the study area inside the bounding box, (b) the Pauli RGB image ($R=\left\langle \left|HH-VV\right|^{2} \right\rangle$, $G=\left\langle \left|2HV\right|^{2} \right\rangle$, $B=\left\langle \left|HH+VV\right|^{2} \right\rangle$) with a horizontal transect 'A' shown in red.}
\label{fig:dataset}
\end{figure}
This particular subset from the UAVSAR image over Hayward is chosen because it consists of clustered areas with varying orientation angles. The radar LOS and flight direction is marked with an arrow in Figure~\ref{fig:dataset}(a) and the subset of the scene is shown with a black border. As seen from the street map, there are variations in the orientation angle of the urban structures with respect to the radar LOS. The top part of the scene consists of an urban patch which is approximately $15^\circ$ to $20^\circ$ oriented away from the radar LOS. The central portion of the scene consists of non-oriented urban areas, while in the lower half of the scene, the urban areas are approximately $-5^\circ$ to $-10^\circ$ oriented away from the radar LOS. The right portion of the scene consists of a forested area over an undulating topography and the left portion consists of a water area. The Pauli RGB image in Figure~\ref{fig:dataset}(b) clearly shows the dominance of $\mbox{HV}$ component (in green) in the urban area rotated along the LOS. 
\section{Results}
\begin{figure}[h!]
\centering
\subfigure[]{\includegraphics[width=0.36\textwidth]{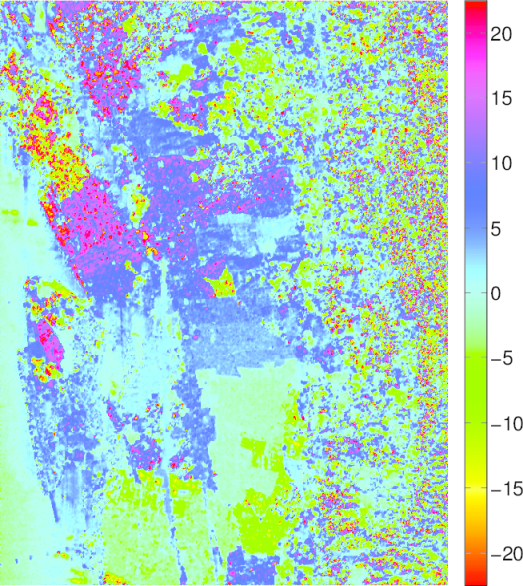}} \hspace{5mm}
\subfigure[]{\includegraphics[width=0.36\textwidth]{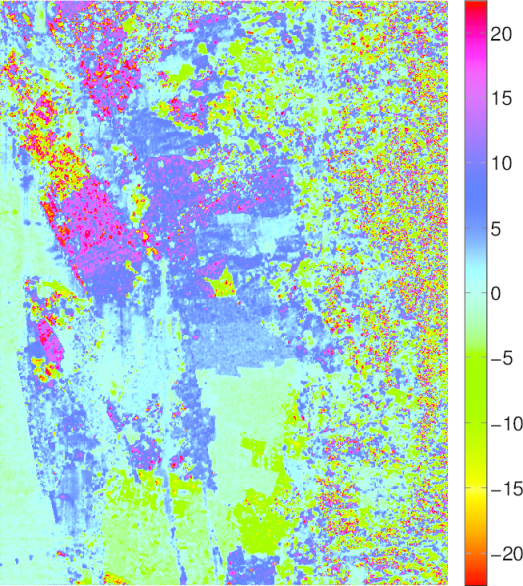}}
\subfigure[]{\includegraphics[width=0.495\textwidth]{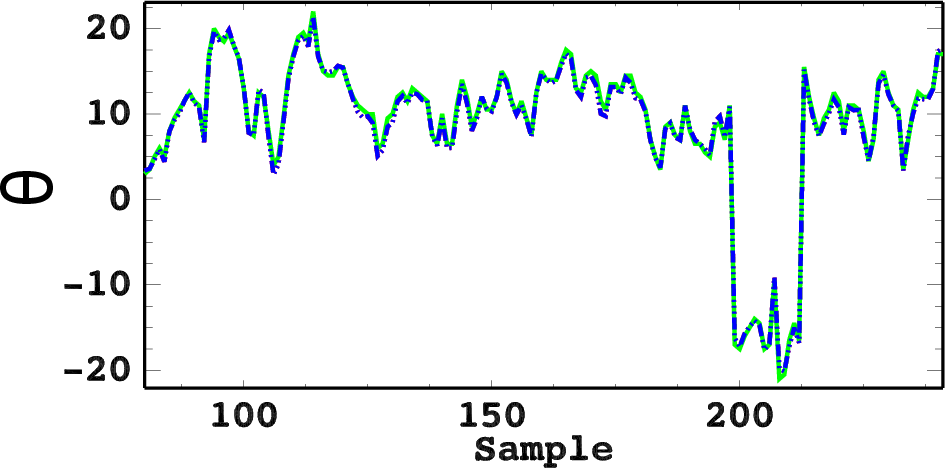}} 
\subfigure[]{\includegraphics[width=0.495\textwidth]{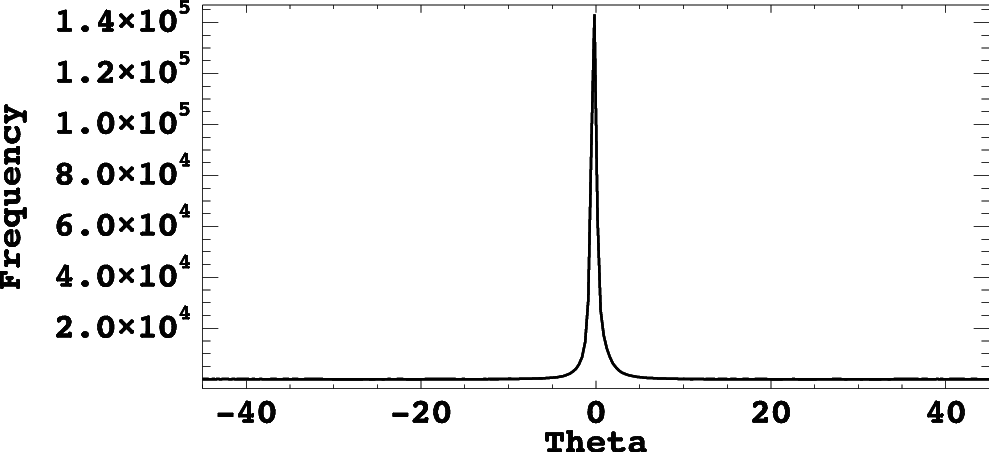}}
\caption{Real orientation angle (a) real orientation angle obtained by the proposed method, (b) real orientation angle obtained by the method of~\citep{Lee2011}, (c) comparison of the estimated orientation angle for the given transect, (d) histogram of the difference in orientation angle computed for the entire scene by the two methods.}
\label{fig:results_ROA}
\end{figure}

\begin{figure}[h!]
\centering
\subfigure[]{\includegraphics[width=0.36\textwidth]{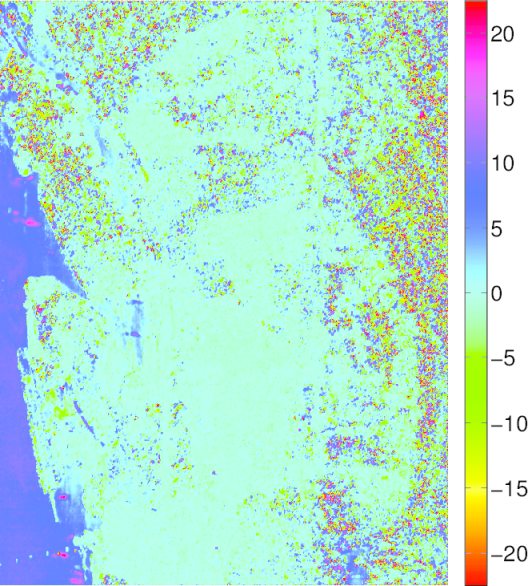}}\hspace{5mm}
\subfigure[]{\includegraphics[width=0.36\textwidth]{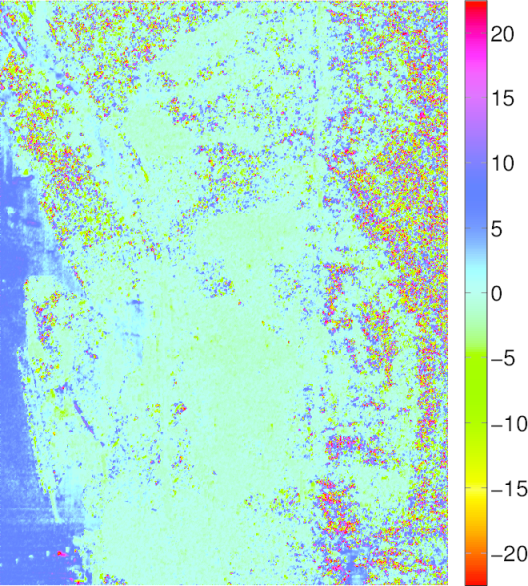}}
\subfigure[]{\includegraphics[width=0.495\textwidth]{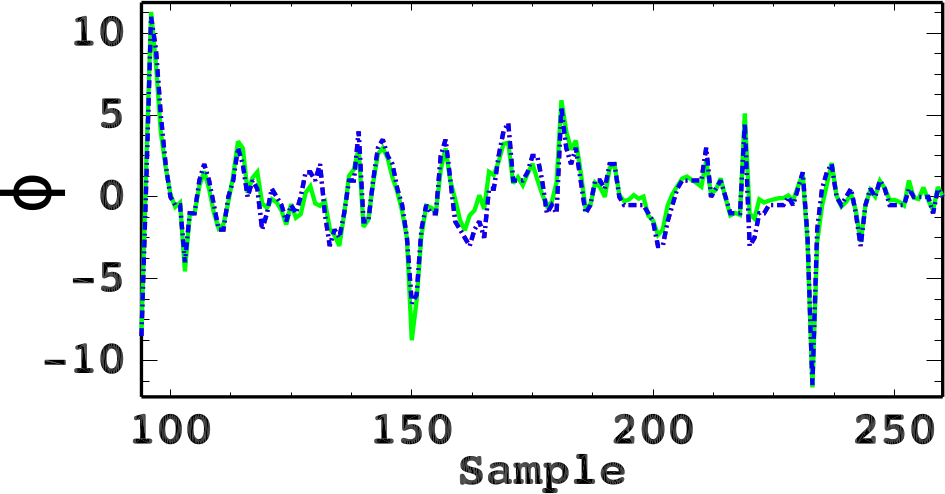}}
\subfigure[]{\includegraphics[width=0.495\textwidth]{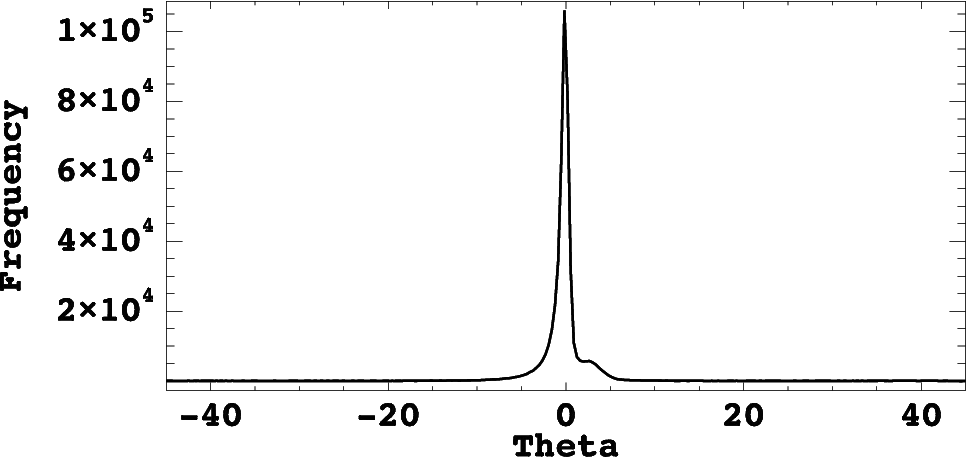}}
\caption{Complex orientation angle (a) complex orientation angle obtained by the proposed method, (b) complex orientation angle obtained by the method of~\citep{Lee2011}, (c) comparison of the estimated orientation angle for the given transect, (d) histogram of the difference in orientation angle computed for the entire scene by the two methods.}
\label{fig:results_ROC}
\end{figure}

The real orientation angle obtained by the proposed method and the one given in~\citep{Lee2011} for the study area (Figure~\ref{fig:dataset}(a)) is shown in Figure~\ref{fig:results_ROA}(a) and Figure~\ref{fig:results_ROA}(b) respectively. 
A 3$\times$3 boxcar filter has been applied to the PolSAR data before the orientation angle estimation. The orientation angle estimated by the proposed method shows complete similarity with that given in Lee~\emph{et~al.}. This is illustrated in Figure~\ref{fig:results_ROA}(c) for a transect "A" (shown in Figure~\ref{fig:dataset}(b)) over a rotated urban area. The histogram of the difference in orientation angle computed for the entire scene by the two methods shows a mean of 0.06$^\circ$ with a standard deviation of 4.2$^\circ$ as shown in Figure~\ref{fig:results_ROA}(d). 
The orientation angle obtained from complex rotation by the method in~\citep{singh13} and the proposed method are shown in Figure~\ref{fig:results_ROC}(a) and Figure~\ref{fig:results_ROC}(b) respectively. The complex orientation angle roughly varies from 0$^\circ-$ 2$^\circ$ for most areas of the image, with the exception of the water surface which shows an orientation angle of about 7$^\circ$. The histogram of the difference in orientation angle computed for the entire scene by the two methods shows a mean of $-$0.04$^\circ$ with a standard deviation of 4.3$^\circ$ as shown in Figure~\ref{fig:results_ROC}(d). 

\begin{figure}[h!]
\centering
\subfigure[]{\includegraphics[width=0.36\textwidth]{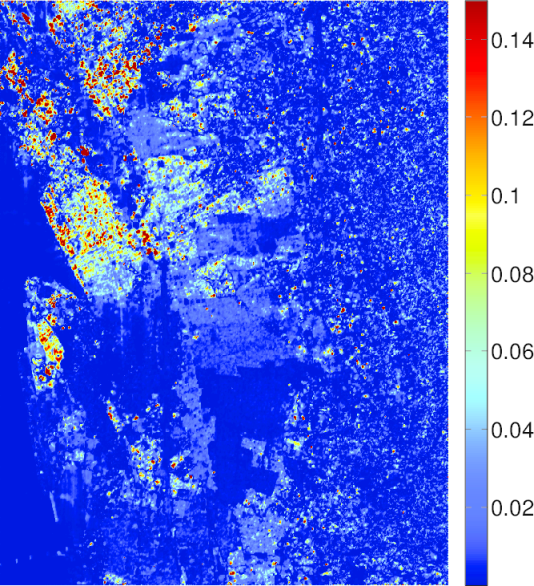}} \hspace{5mm}
\subfigure[]{\includegraphics[width=0.36\textwidth]{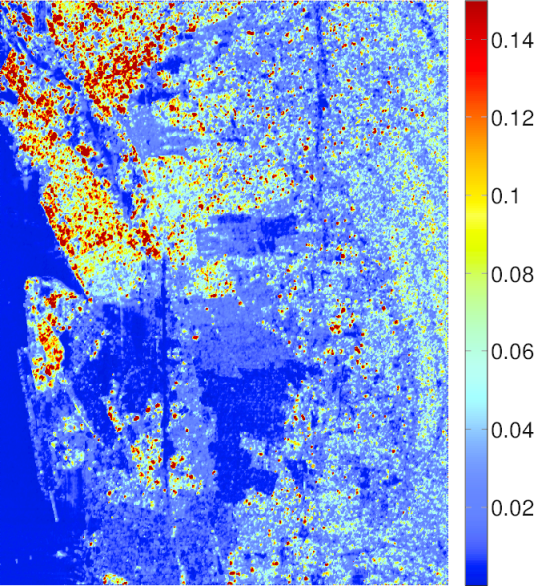}}
\subfigure[]{\includegraphics[width=0.6\textwidth]{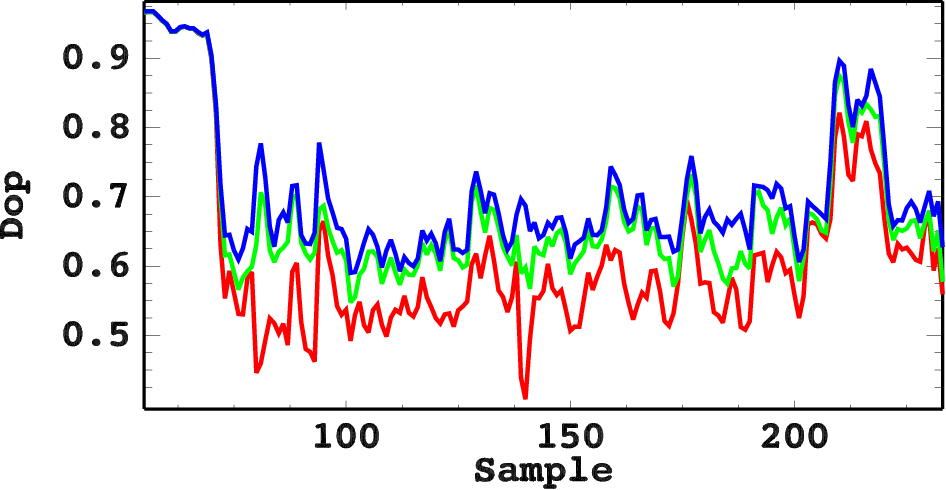}}
\caption{The variation of the degree of polarization (Dop) with orientation (a) $\Delta p_{E}=p_{E}(\theta_{0})-p_E$, (b) $\Delta p_{E}^{'}=p_{E}(\phi_{0})-p_E$ (UR - Un Rotated, RR - Real Rotated, CR - Complex Rotated).}
\label{fig:dop_change_R_C}
\end{figure}

The Dop after real orientation compensation will either increase or remain the same as the unrotated case, and the Dop after complex orientation compensation will subsequently either increase or remain the same as the real rotated case. Figure~\ref{fig:dop_change_R_C}(a) shows the change in Dop between the real rotated and the unrotated image, while Figure~\ref{fig:dop_change_R_C}(b) depicts the Dop change between the complex rotated and unrotated image. It can be seen that for a urban patch in the top-left portion of the image, the Dop increases by $\sim8\%$ for the real rotation case, whereas the same area shows an increase of $\sim12\%$ in case of the complex rotation. This is illustrated by the transect drawn over the urban patch, shown in Figure~\ref{fig:dop_change_R_C}(c). A similar rise of approximately $\sim2\%$ is also seen in the forested areas growing over undulating terrain in the right portion of the image. There is negligible or no rise in Dop observed for the water area that dominates the left portion of the image. This is shown in Figure~\ref{fig:dop_change_R_C}(c), where the portion of the transect to the left of the dotted line represents the water surface.  This interesting insight into the variation of Dop with real and complex orientation compensation will be useful in characterizing a target. 
\section{Conclusion}
An orientation compensation method for PolSAR data has been developed and analyzed with a L-band UAVSAR data. We have found that the angle extracted from the proposed method is similar to the one proposed by an earlier method. An insight into the effect of orientation compensation is revealed by the variation of Dop with real and complex orientation. For a completely polarized incident wave, the output Dop is dependent on the Mueller matrix which describes the scattering process. In this regard, further research is needed to estimate the maximum achievable Dop for a target under consideration. 

\bibliographystyle{tRSL}
\bibliography{mybibfile}

\end{document}